\documentclass[12pt]{article}
\usepackage{amsmath}
\usepackage{amssymb}
\usepackage[dvips]{graphicx}
\usepackage[dvips]{epsfig}
\usepackage{color}

 \advance \textwidth by 1cm
\def\##1{{\bf #1}}
\def\=#1{\underline{\underline #1}}

\def\.{\mbox{ \tiny{$^\bullet$} }}

\def\le{\left(}
\def\ri{\right)}
\def\les{\left[}
\def\ris{\right]}

\def\ric{\right\}}

\def\c#1{\cite{#1}}

\def\epso{\varepsilon_{\scriptscriptstyle 0}}
\def\lambdao{\lambda_{\scriptscriptstyle 0}}

\def\ko{k_{\scriptscriptstyle 0}}
\def\co{c_{\scriptscriptstyle 0}}

\def\eps{\varepsilon}
\def\epsa{\eps_a}
\def\epsb{\eps_b}
\def\epsc{\eps_c}

\def\epsr{{\underline{\underline \eps}}_{ani}}
\def\epsiso{\eps_{iso}}
 
 \def\dprop{\Delta_{prop}}
 \def\vph{v_{ph}}

\def\ux{\hat{\#u}_x}
\def\uy{\hat{\#u}_y}
\def\uz{\hat{\#u}_z}

\begin{document}

\begin{center}
\textbf{Dyakonov--Tamm waves guided jointly by an ordinary, isotropic, homogeneous, dielectric material and a hyperbolic, dielectric, structurally chiral material }\\
\vspace{0.5cm}

{Akhlesh Lakhtakia and Muhammad Faryad  }\\

\vspace{0.5cm}

{ Pennsylvania State University, Department of Engineering Science and
Mechanics, Nanoengineered Metamaterials Group (NanoMM),
University Park, PA  16802--6812, USA\\
}
\vspace{0.5cm}

 \end{center}

\pagestyle{plain} 
 
\begin{abstract}
The planar interface of an ordinary, isotropic, homogeneous, dielectric material and a hyperbolic, dielectric, structurally chiral material can support the propagation of  one or multiple Dyakonov--Tamm waves, at a specified frequency and along a specified direction in the interface plane. When multiple Dyakonov--Tamm  waves can exist, they differ in phase speed, propagation length, degree of localization to the interface, and spatial profiles of the associated electromagnetic fields. Dependence on the relative permittivity scalar of the isotropic partnering material suggests exploitation for optical sensing of analytes.
 \end{abstract}

\section{Introduction}
The Dyakonov wave is an electromagnetic surface wave
guided by the planar interface of two homogenous dielectric materials, at least one of which is anisotropic. The existence of the Dyakonov wave was predicted in the 1980s \cite{Marchevskii,Dyakonov88} and experimentally verified about five years ago \cite{Takayama09}. The 
Tamm wave is  guided by the planar interface of
two isotropic dielectric materials, at least one of which is periodically 
nonhomogeneous in the direction normal to the interface. Predicted in 1977 \cite{YYH}, Tamm waves
were experimentally observed shortly thereafter \cite{YYC} and are being applied for optical sensing of
analytes \cite{SR,KA}. 

Whereas the anisotropy of a partnering material provides the Dyakonov wave a  sensitivity to the direction of propagation
\cite{Takayama08},
the periodic nonhomogeneity of a partnering material is responsible for multiple Tamm waves to propagate in any direction \cite{Maab}. Both of these attributes are combined in 
Dyakonov--Tamm waves, whose propagation is guided by the interface of two dielectric materials, one of which must be anisotropic and periodically nonhomogeneous normal to the interface~\cite{LP2007,FLpra2011}.
Recently, the existence of these Dyakonov--Tamm waves was confirmed experimentally  using a prism-coupled configuration~\cite{Pulsifer_13_2}.

In the foregoing papers on Dyakonov--Tamm waves, one partnering material is an ordinary, isotropic, 
homogeneous, dielectric material with a
purely real relative permittivity $\epsiso>0$, whereas the real part of the relative permittivity dyadic $\epsr$  of the anisotropic, periodically nonhomogeneous, dielectric  partnernering material is positive definite, i.e., all three eigenvalues of $\epsr$ have positive real parts \cite{Matrices}. What would happen if ${\rm Re}(\epsr)$ were indefinite, i.e., if either one or two of its eigenvalues had negative real parts, the remaining having positive real parts? Such materials are nowadays called hyperbolic materials. Homogeneous {hyperbolic materials}  exist in nature \cite{FG,Gerbaux,Sun} and have also been manufactured
 \cite{Kanungo,Othman}.  Periodically nonhomogeneous {hyperbolic materials} appear very likely to be manufacturable \cite{Lakhjnp2014}
 using   physical-vapor-deposition techniques \cite{Hawkeye,Raulbook}.

In this Letter, we {present the results of our investigations on} Dyakonov--Tamm waves guided by the planar interface of an ordinary, isotropic, homogenous, dielectric material and a hyperbolic, dielectric, structurally chiral material which is nonhomogeneous in the direction normal to the interface.
Section \ref{theory} briefly
describes
the relevant one-point boundary-value problem. Numerical results are provided and discussed
in Sec.~\ref{nrd}.  An $\exp(-i{\omega}t)$ dependence on time $t$ is implicit, with $i=\sqrt{-1}$ and
$\omega$ denoting the angular frequency.

\section{Theoretical Preliminaries}\label{theory}
The one-point boundary-value problem of Dyakonov--Tamm-wave propagation is as follows:
Suppose that the plane $z=0$ is the interface between the two chosen partnering materials. The  half space $z<0$
is occupied by an  ordinary, isotropic, homogenous, dielectric {material} with relative permittivity $\epsiso$ such that
${\rm Re}(\epsiso)>0$ and ${\rm Im}(\epsiso)=0$. The half space $z>0$ is occupied by
a hyperbolic, dielectric, structurally chiral material with relative permittivity dyadic \cite{Lakhjnp2014}
 \begin{equation}
\epsr(z)=\epso\, \=S_z(z)\.
 \=S_y(\chi)\. \left(
\epsa  \, \uz\uz+ \epsb \, \ux\ux
 +\, \epsc \, \uy\uy\right)
\.\=S_y^{-1}(\chi)\.\=S_z^{-1}(z)\,,
\end{equation}
where the direction of nonhomogeneity is parallel to the
$z$ axis; the Cartesian unit vectors are
identified as $\ux$, $\uy$, and $\uz$;   and $\epso$ is  the  permittivity of
free space; the periodic nonhomogeneity  is expressed through the
rotation dyadic
\begin{equation}
\=S_z(z)=\uz\uz +\le\ux\ux+\uy\uy\ri\cos\le\frac {h\pi z}{\Omega}\ri
+\le\uy\ux-\ux\uy\ri\sin\le\frac {h\pi z}{\Omega}\ri\,,
\label{Sz-def}
\end{equation}
with $2\Omega$ as the period and either $h=+1$ for structural right-handedness
or $h=-1$ for structural left-handedness; the dyadic
\begin{equation}
\=S_y(\chi)=\le \ux\ux + \uz\uz \ri \cos{\chi}
+\le \uz\ux -
\ux\uz \ri \sin{\chi}+\uy\uy 
\label{Sy-def}
\end{equation}
contains the tilt angle $\chi\in[0,\pi/2]$ with respect to the $xy$ plane; 
$\epsa$, $\epsb$, and $\epsc$ are the three $z$-independent eigenvalues of  $\epsr(z)$;
and either one or two of these three eigenvalues have negative real parts but the remainder do not.

The   electromagnetic field phasors everywhere can be written as \cite{LP2007}
\begin{equation}
\left.\begin{array}{l}
\#E(\#r)=\#e(z) \exp\les{i}q(x\cos\psi+y\sin\psi)\ris\\[5pt]
 \#H(\#r)=\#h(z) \exp\les{i}q(x\cos\psi+y\sin\psi)\ris
 \end{array}\ric\,,
 \end{equation}
 with the unknown surface wavenumber $q$ and unknown functions $\#e(z)$ and $\#h(z)$. The angle $\psi\in[0,2\pi)$ denotes the direction
 of propagation in the $xy$ plane. As the procedure to obtain and solve a dispersion equation for $q$, 
 and then determine $\#e(z)$ and $\#h(z)$, for a specific $\psi$ has been described elsewhere
 in detail \cite{LP2007}, it is not repeated here.

 \section{Numerical results and discussion}\label{nrd}

 For illustrative results, we set $\epsa=2.26(1+i\delta)$,
 $\epsb=3.46(-1+i\delta)$, $\epsc=2.78(1+i\delta)$, $\delta=0.001$, $h=+1$, $\Omega=135$~nm, $\chi=\pi/6$, and $\psi=0$. 
 Without loss of generality, we took the isotropic partner to be free space (i.e., $\epsiso=1$) \c{FLP}. Our search was mostly restricted to the
 regime ${\rm Re}(q)/\ko\in(1,4]$, where $\ko$ is the free-space wavenumber.

\begin{figure}[ht]
\begin{center}
\includegraphics[width=10cm]{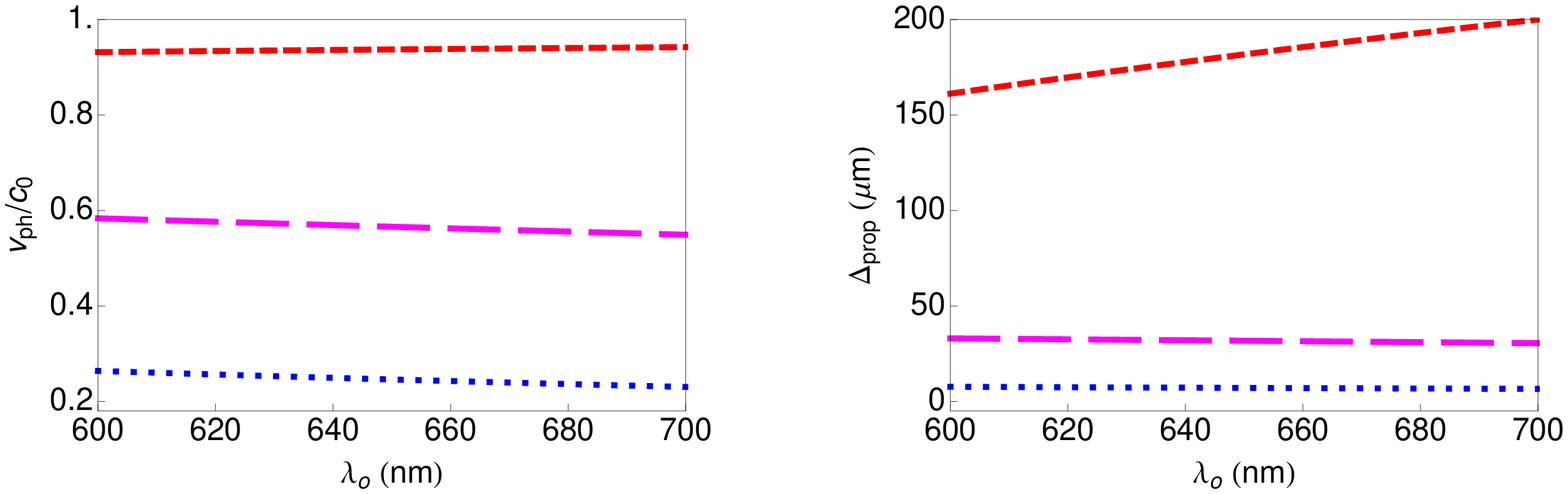}
\caption{Phase speed $\vph$ and propagation length
$\dprop$ as functions of $\lambdao$, when $\epsiso=1$, $\epsa=2.26(1+i\delta)$,
 $\epsb=3.46(-1+i\delta)$, $\epsc=2.78(1+i\delta)$, $\delta=0.001$, $h=+1$, $\Omega=135$~nm, $\chi=\pi/6$, and $\psi=0$. 
 The solutions of the dispersion equation are organized in three branches.
 }%
\label{fig1}
\end{center}
\end{figure}

 At every free-space wavelength
 $\lambdao\in[600,700]$~nm, we found   three solutions $q$ of the  dispersion equation.  The solutions were organized
 in three branches,
 as shown in Fig.~\ref{fig1}. Thus, three distinct Dyakonov--Tamm
 waves  differing in phase speed $\vph=\omega/{\rm Re}(q)$ and propagation length $\dprop=1/{\rm Im}(q)$ can propagate along the
 $x$ axis in the interface plane for $\lambdao\in[600,700]$~nm. Dyakonov--Tamm waves on the highest-$\vph$ branch have $\vph \simeq0.935\co$, where $\co=\omega/\ko$ is the speed of light in free space, and
 $\dprop$ ranging from $\sim160$~nm to $\sim200$~nm.
Dyakonov--Tamm waves on the lowest-$\vph$ branch have $\vph$ ranging from $\sim0.23\co$ to $\sim0.264\co$,  and
 $\dprop\simeq7.0$~$\mu$m. Clearly from the presented data, a lower phase speed is associated with higher attenuation along the direction of propagation.

\begin{figure}[!ht]
\begin{center}
\includegraphics[width=10cm]{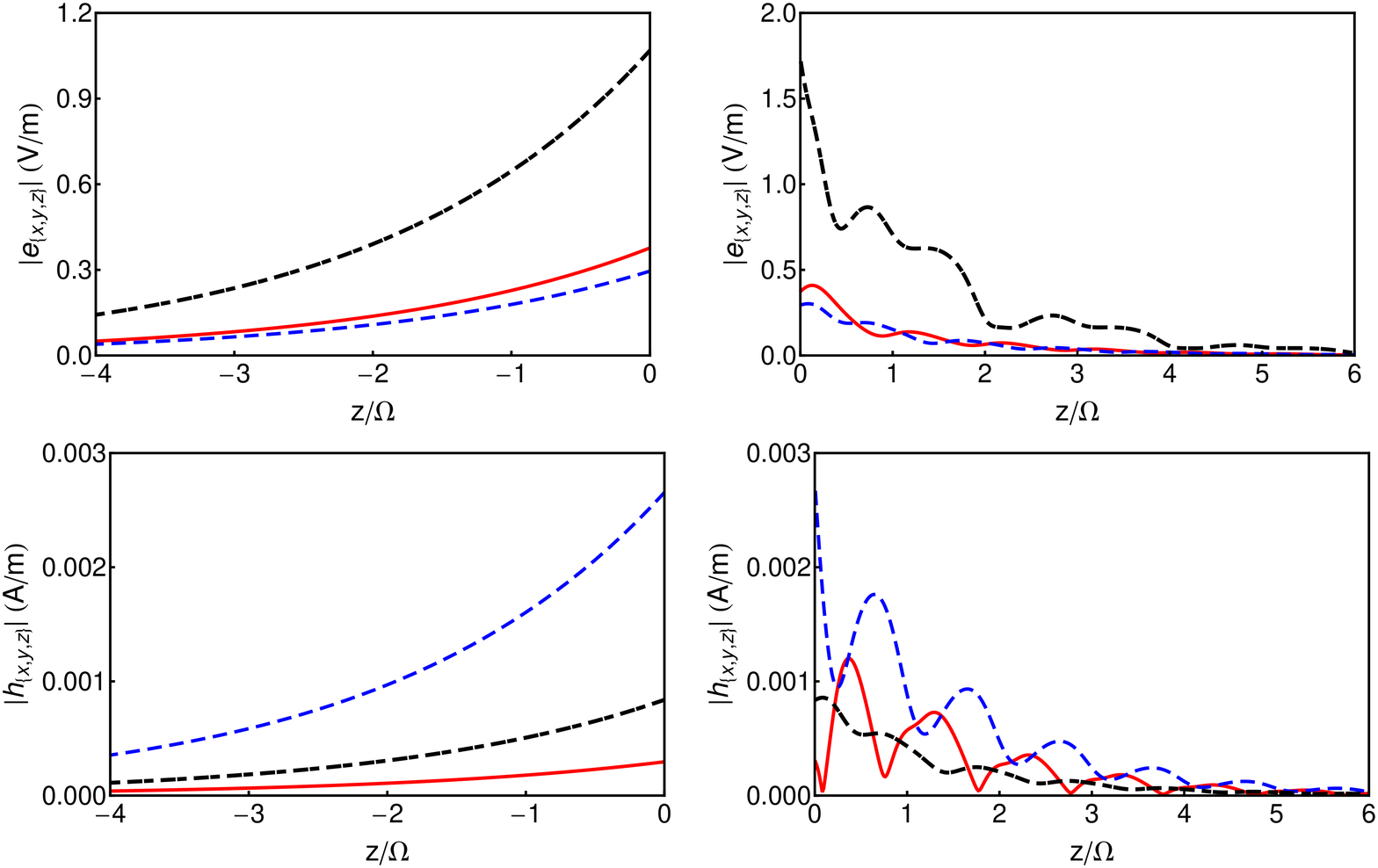}
\caption{(Color online) Normalized variations with $z$ of the magnitudes of the Cartesian components of $\#e(z)$ and $\#h(z)$ of an SPP wave 
when  $\lambdao=635$~nm,  $\epsiso=1$, $\epsa=2.26(1+i\delta)$,
 $\epsb=3.46(-1+i\delta)$, $\epsc=2.78(1+i\delta)$, $\delta=0.001$, $h=+1$, $\Omega=135$~nm, $\chi=\pi/6$, and $\psi=0$. {The $x$, $y$, and $z$-components are represented by red solid, blue dashed, and black chain-dashed lines, respectively.}
For this Dyakonov--Tamm wave,  $q=(1.0686  + i0.0006)\ko$.  
}
\label{fig2}
\end{center}
\end{figure}

\begin{figure}[!ht]
\begin{center}
\includegraphics[width=10cm]{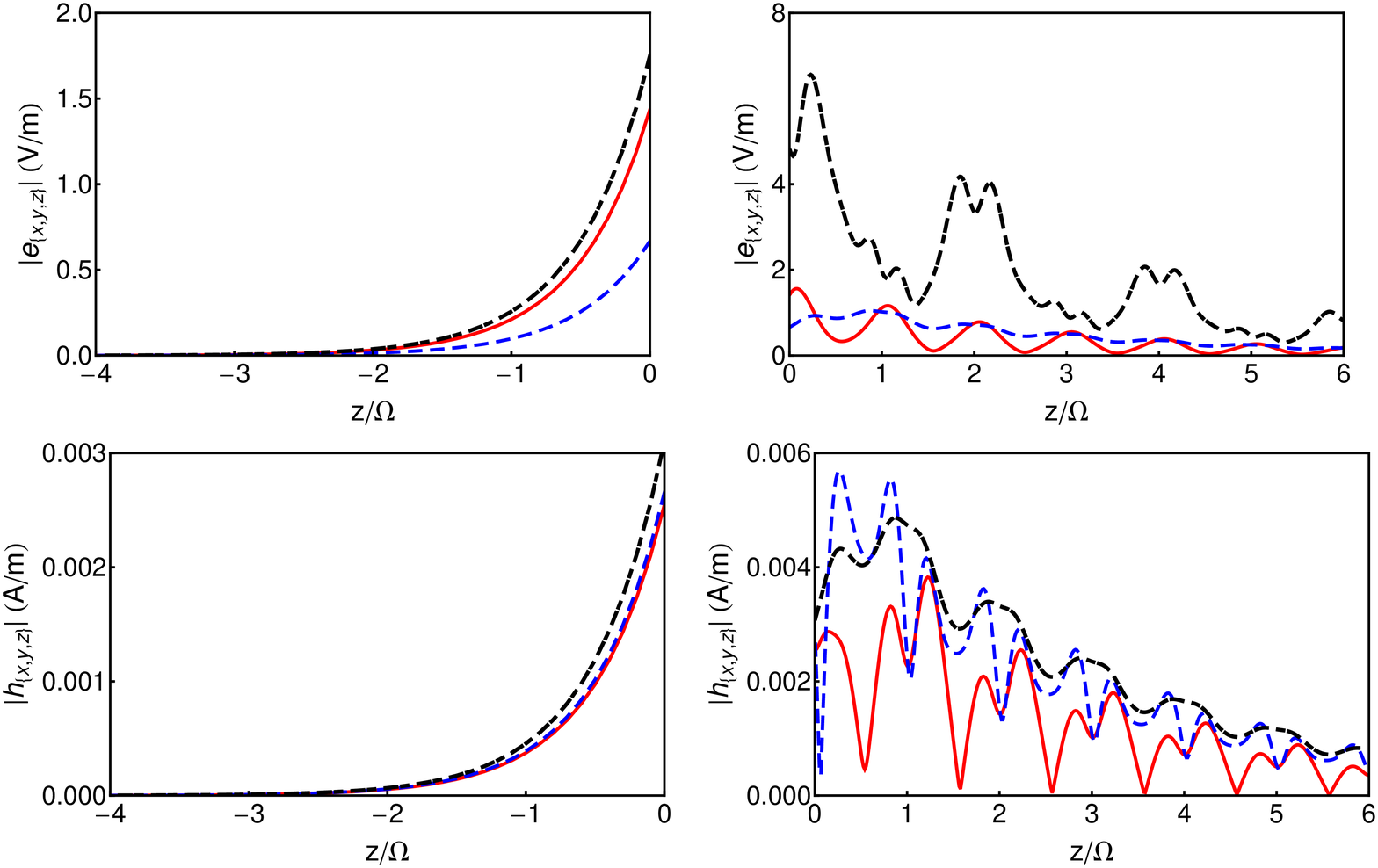}
\caption{(Color online) Same as Fig.~\ref{fig2}, except for the Dyakonov--Tamm wave with $q=(1.7507 + i 0.0031)\ko$.}
\label{fig3}
\end{center}
\end{figure}

\begin{figure}[!ht]
\begin{center}
\includegraphics[width=10cm]{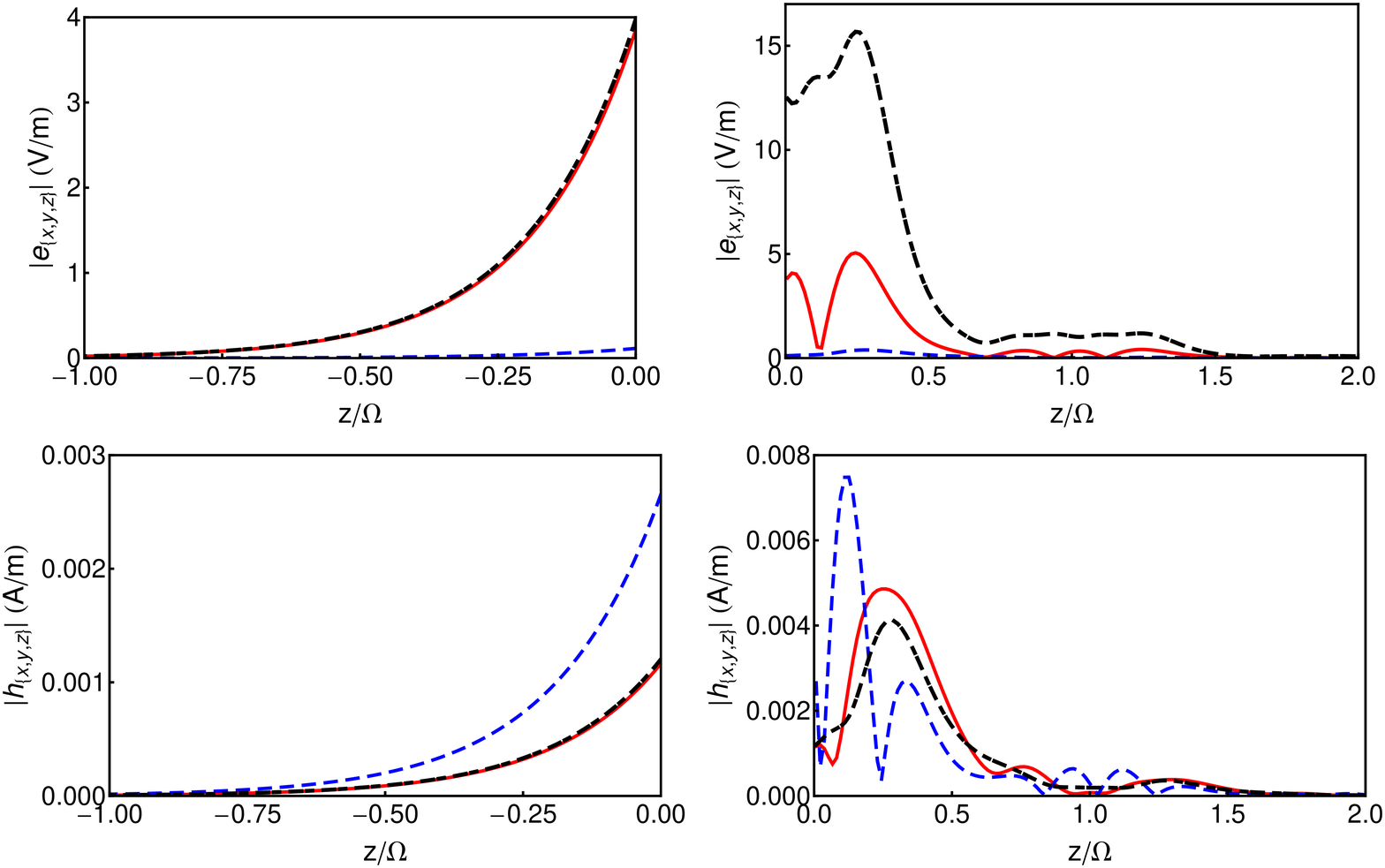}
\caption{(Color online) Same as Fig.~\ref{fig2}, except for the Dyakonov--Tamm wave with $q=(3.9809 + i0.0139)\ko$.}
\label{fig4}
\end{center}
\end{figure}

The Dyakonov--Tamm waves on the three branches  in  Fig.~\ref{fig1} are not only dissimilar in phase speed and propagation length, but the spatial profiles of their field
along the $z$ axis also differ. Let us focus on the three solutions of the dispersion equation 
for $\lambdao=635$~nm. Figure~\ref{fig2} presents plots of the magnitudes of the Cartesian components of $\#e(z)$ and $\#h(z)$ as
functions of $z$ for the Dyakonov--Tamm wave with $q=(1.0686  + i0.0006)\ko$,  Fig.~\ref{fig3}   for the Dyakonov--Tamm
wave with  $q=(1.7507 + i 0.0031)\ko$, and Fig.~\ref{fig4} for  the Dyakonov--Tamm
wave with  $q=(3.9809 + i0.0139)\ko$.   On the isotropic side, the degree of localization of the Dyakonov--Tamm wave
to the interface is greater if  ${\rm Re}(q)$ is higher.   However,
on the anisotropic side of the interface, a clear dependence of localization to the interface does not emerge from the three figures. Indeed, of the three Dyakonov--Tamm waves,
the  one in Fig.~\ref{fig4} is the most strongly localized to the interface, whereas the one in Fig.~\ref{fig3} is  the most  weakly localized to the interface.

\begin{figure}[!ht]
\begin{center}
\includegraphics[width=10cm]{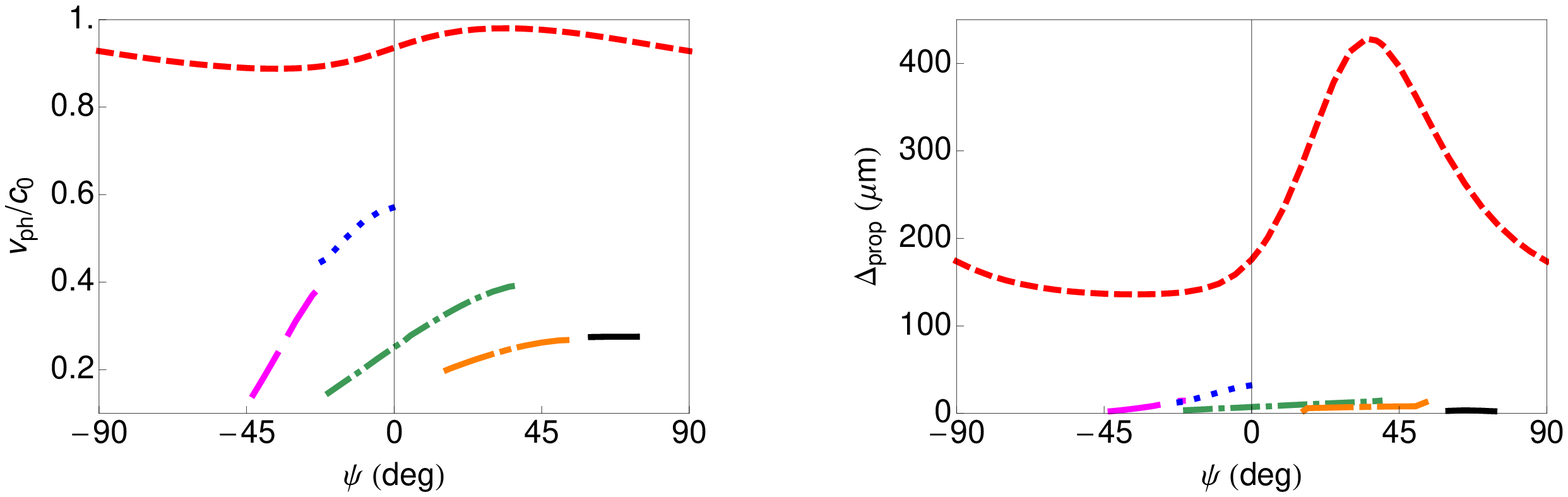}
\caption{(Color online) Phase speed $\vph$ and propagation length
$\dprop$  as functions of $\psi$, when $\lambdao=635$~nm, $\epsiso=1$, $\epsa=2.26(1+i\delta)$,
 $\epsb=3.46(-1+i\delta)$, $\epsc=2.78(1+i\delta)$, $\delta=0.001$, $h=+1$, $\Omega=135$~nm,  and $\chi=30^\circ$. The solutions of the dispersion equation are organized in six branches.}
\label{fig5}
\end{center}
\end{figure}

The  Dyakonov--Tamm-wave-propagation phenomenon depends strongly on $\psi$. This is exemplied by
 the plots in Fig.~\ref{fig5} of $\vph$  and $\dprop$   as functions of $\psi\in\left[-90^\circ,90^\circ\right]$ when $\lambdao=635$~nm and $\chi=30^\circ$. The solutions can be organized in six branches. One branch, with the highest values of $\vph$ and $\dprop$ spans the entire angular regime $\psi\in\left[-90^\circ,90^\circ\right]$ available for propagation. Each of the remaining five branches  spans a finite range of $\psi$. Although the  five branches appear to be clustered together,
 the phase speeds range from $0.14\co$ to $0.58\co$
 and the propagation lengths from $\sim2$~$\mu$m to $\sim34$~$\mu$m, which are quite wide ranges. Either one or two or three Dyakonov--Tamm waves can propagate in the interface plane at an angle $\psi$ with respect to the $x$ axis.
 Let us also note that, because of chiral symmetry about the $z$ axis, the solutions  of the dispersion equation are the same for $\psi$ and $\psi\pm180^\circ$. 
 
 The  Dyakonov--Tamm-wave-propagation phenomenon also depends strongly on $\epsiso$. Suppose that free space, the isotropic partnering material for Figs.~\ref{fig2}--\ref{fig4}, were to be replaced by a dielectric material
with $\epsiso=2.25$. Then the   dispersion equation has just only two solutions---$q=(1.9227 + i0.0040)\ko$ and $q=(4.2766 +i0.0150)\ko$---instead of  the three solutions for $\epsiso=1$. The large changes between the solutions for the two  values of $\epsiso$ chosen here strongly suggest the potential of Dyakonov--Tamm waves supported by 
a partnering material that is hyperbolic, dielectric, and structurally chiral for optical sensing of analytes in a prism-coupled configuration, just as both surface-plasmon-polariton waves~\cite{Homola} and Tamm waves~\cite{SR,KA} are being used
by technoscientists.

\bigskip
\noindent {\bf Acknowledgments.}
 A.L. is grateful to the Charles Godfrey Binder Endowment at
the Pennsylvania State University for the financial support of
his ongoing research. M.F. was supported by Grant
No. DMR-1125591 from the U.S. National Science Foundation.

\end{document}